\documentclass[aip,reprint,floatfix,superscriptaddress]{revtex4-2}

\usepackage{amsmath}
\usepackage{amssymb}
\usepackage{accents}
\usepackage{tabularx}
\usepackage{graphicx}
\usepackage{floatrow}
\usepackage{rotating}
\usepackage{dcolumn}
\usepackage{xr}
\usepackage{cleveref}
\usepackage{bm}
\usepackage{xcolor}
\UseRawInputEncoding

\usepackage{verbatim}

\begin{document}

\date{\today}\title{Mode-Dependent Scaling of Nonlinearity and Linear Dynamic Range in a NEMS Resonator}

\newcommand{\BU}{Department of Mechanical Engineering, Division of Materials Science and Engineering, and the Photonics Center, Boston University, Boston, Massachusetts 02215, United States}

\newcommand{\VT}{Department of Mechanical Engineering, Virginia Tech, Blacksburg, Virginia 24061, United States}

\newcommand{\GC}{Department of Physics, Gordon College, Wenham, Massachusetts 01984, United States}

\newcommand{\SUNUM}{SUNUM, Nanotechnology Research and Application Center, Sabanci University, Istanbul, 34956, Turkey}

\newcommand{\Sabanci}{Faculty of Engineering and Natural Sciences, Sabanci University, Istanbul, 34956, Turkey}

\newcommand{\Bilkent}{Department of Mechanical Engineering, Bilkent University, Ankara, 06800, Turkey}

\newcommand{\UNAM}{National Nanotechnology Research Center (UNAM), Bilkent University, Ankara, 06800, Turkey}

\author{M. Ma}
%\email[Electronic mail: ]{monan@bu.edu}
\affiliation{\BU}

\author{N. Welles}
\affiliation{\VT}

\author{O. Svitelskiy}
\affiliation{\GC}

\author{C. Yanik}
\affiliation{\SUNUM}

\author{I. I. Kaya}
\affiliation{\SUNUM}
\affiliation{\Sabanci}

\author{M. S. Hanay}
\affiliation{\Bilkent}
\affiliation{\UNAM}

\author{M. R. Paul}
\affiliation{\VT}

\author{K. L. Ekinci}
\email[Electronic mail: ]{ekinci@bu.edu}
\affiliation{\BU}

\date{\today}

\begin{abstract}
Even a relatively weak drive force is enough to push a typical nanomechanical resonator into the nonlinear regime. Consequently, nonlinearities are widespread in nanomechanics and determine the critical characteristics of nanoelectromechanical systems (NEMS) resonators. A thorough understanding of the nonlinear dynamics of higher eigenmodes of NEMS resonators would be beneficial for progress, given their use in applications and fundamental studies. Here, we characterize the  nonlinearity and the linear dynamic range (LDR) of each  eigenmode of two nanomechanical beam resonators with different intrinsic tension values up to eigenmode $n=11$. We find that the modal Duffing constant increases as $n^4$, while the critical amplitude for the onset of nonlinearity decreases as $1/n$. The LDR, determined from the ratio of the critical amplitude to the thermal noise amplitude, increases weakly with $n$. Our findings are consistent with our theory treating the beam as a string, with the nonlinearity emerging from stretching at high amplitudes. These scaling laws, observed in experiments and validated theoretically, can be leveraged for pushing the limits of NEMS-based sensing even further.  
\end{abstract}

\maketitle

Over the past decade, the sensitivities attainable by nanoelectromechanical systems (NEMS) have steadily approached fundamental limits.\cite{moser2013ultrasensitive,nan2013self,hanay2015inertial,heritier2018nanoladder,de2018ultrasensitive,fogliano2021ultrasensitive} In a typical NEMS sensing application, one drives the NEMS resonator at one of its resonances and looks for changes in its resonance frequency or  amplitude due to a prescribed interaction with the environment. The highest sensitivity  is attained when the mode is driven to the largest  amplitude possible and its oscillations are detected by the lowest-noise motion transducer available. \cite{ekinci2004ultrasensitive} All these considerations are captured in a widely-used and intuitive formula that provides the minimum detectable frequency shift (\emph{i.e.}, rms frequency noise) $\delta f$ for a NEMS-based resonant sensor:\cite{ekinci2004ultimate}
\begin{equation}\label{eqn:sensitivity}
\delta f \approx {\frac{f}{2Q}} \, {10^{ -\frac{\rm{LDR}}{20}}}.
\end{equation}
Here, $f$ and $Q$ are the resonance frequency and the quality factor of the NEMS mode, respectively; LDR (in units of dB) is the linear dynamic range of the mode defined as\cite{postma2005dynamic}
\begin{equation}\label{eqn:dr_initial}
\mathrm{LDR}  = 20\log{\left(\frac{0.745z_{c}}{z_N}\right)}.
\end{equation}
In Eq. (\ref{eqn:dr_initial}), $z_c$  is the so-called critical amplitude (rms) for the onset of nonlinearity of the mode: more precisely, the  frequency-response curve for the mode first attains infinite slope at $z_c$  as the drive force is increased;  $z_N$ is the rms noise amplitude within the measurement bandwidth. The numerical factor quantifies the 1 dB compression that occurs when the signal is 1 dB below the expected linear response. It is immediately evident from Eqs. (\ref{eqn:sensitivity}) and (\ref{eqn:dr_initial}) that, to maximize the signal-to-noise ratio (SNR) in \textit{linear} NEMS operation, the NEMS resonator must be driven to the cusp of nonlinearity.\cite{roy2018improving, demir2019fundamental} A  formula similar to Eq. (\ref{eqn:sensitivity}) exists for amplitude detection. \cite{martin1987atomic}  

\begin{figure*}
    \floatbox[{\capbeside\thisfloatsetup{capbesideposition={right,top},capbesidewidth=0.3\textwidth}}]{figure}[\FBwidth]
{\caption{(a) SEM image of a NEMS doubly clamped beam with thin-film gold nanoresistors  around both anchors. (b) Experimental and theoretical eigenfrequencies $\omega_n/2\pi$ shown on a semilogarithmic plot. The upper inset is a numerical simulation of the $7^{\rm th}$ eigenmode; the lower inset shows $\omega_n/\omega_1$ on a linear plot. (c) Experimental and theoretical spring constants $k_n$ shown on a semilogarithmic plot. Error bars are smaller than the symbols unless explicitly shown. Inset shows the PSD of the thermal fluctuations of the $7^{\rm th}$ mode with a Lorentzian fit to the peak. (d) Experimental quality factors $Q_n$ decrease as a function of $n$. Inset shows the same data on a semilogarithmic plot.}
\label{exposition_linear}}
{\includegraphics[width=0.65\textwidth]{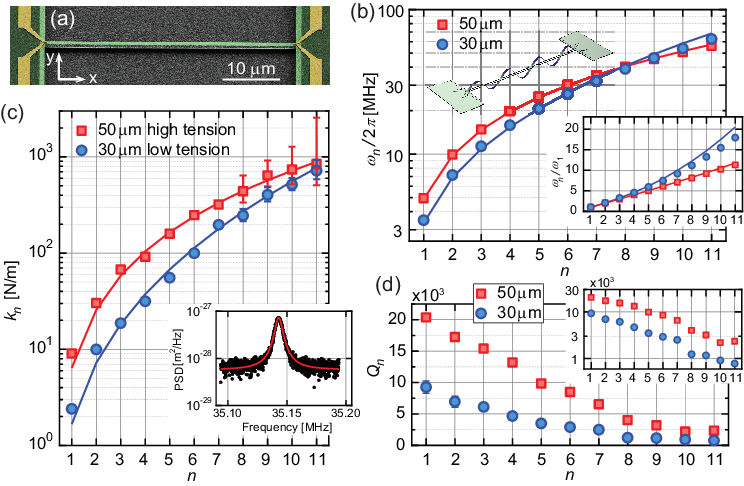}}
\end{figure*}

The above discussion suggests that it is  important to understand and characterize  nonlinearities in NEMS. In particular, the LDR---and, consequently, the onset of nonlinearity and the amplitude of thermal fluctuations---of a NEMS resonator is one of the most essential device parameters for sensing. As a NEMS resonator is uniformly scaled down, its spring constant tends to become smaller and thermal fluctuations become more prominent. \cite{gress2023multimode} This indicates that the LDR will shrink from the bottom end. Experiments on different structures \cite{husain2003nanowire,postma2005dynamic,parmar2015dynamic,molina2021high} have shown that the amplitude for onset of nonlinearity  becomes progressively smaller for smaller structures, suggesting that the top end of the LDR also gets diminished.  In other words, as NEMS get miniaturized,  the LDR tends to keep shrinking,\cite{kozinsky2006tuning,cho2010tunable} leading to next-generation devices that practically have no linear regime.\cite{gieseler2013thermal, barnard2019real}

The quintessential NEMS resonator, in which intrinsic nonlinearities have been carefully explored, is a doubly clamped nanomechanical beam vibrating in its fundamental mode. \cite{postma2005dynamic,kozinsky2007basins,cho2010tunable,huber2020spectral} The most commonly studied intrinsic nonlinearity emerges through tension that develops as the beam stretches at high amplitudes. Analytical description of this system starting with the elastic beam equation leads to the Duffing equation for the fundamental mode. \cite{lifshitz2008nonlinear,nayfeh2008nonlinear,landau2013course,bachtold2022mesoscopic} A quadratic term may also emerge if a broken symmetry exists in the system.\cite{ochs2021resonant,keskekler2022symmetry,kozinsky2006tuning} The analytical theory accurately predicts both the Duffing constant and the onset of nonlinearity for the fundamental mode of NEMS beams made out of different materials\cite{younis2003study,kozinsky2007basins,moskovtsev2017strong,samanta2018tuning} and with different intrinsic tension levels.\cite{Castellanos2013single,cho2010tunable,hajjaj2018two,ochs2022frequency} This consequently allows for a reasonable estimate of the LDR.\cite{postma2005dynamic,kozinsky2006tuning}

As NEMS sensor work progresses, it has become more common to employ higher modes or multiple modes  in applications.\cite{erdogan2022atmospheric, malvar2016mass} With some notable exceptions,\cite{lifshitz2008nonlinear, truitt2013linear, maillet2018measuring, kozinsky2007nonlinear} there has not been much insight into the nonlinearities and LDR of high modes. In this manuscript, we provide  a theoretical framework and experimental measurements  that describe how intrinsic nonlinearity scales with eigenmode number $n$ in a NEMS resonator. We also measure the thermal noise of each eigenmode, and elucidate the $n$ dependence of the LDR. We conclude that the useful LDR increases with increasing intrinsic tension in the device and  $n$, albeit slowly.

We begin with the Euler-Bernoulli beam equation with  an intrinsic tension term. We also include a nonlinear tension term resulting from the stretching of the beam that occurs for large amplitude oscillations. This can be expressed as \cite{postma2005dynamic,kozinsky2007basins,lifshitz2008nonlinear} \begin{multline}\label{beam}
    \rho A{{{\partial ^2} Z(x,t)} \over {\partial {t^2}}} + EI{{{\partial ^4}Z(x,t) } \over {\partial {x^4}}}\\ - \left[ F_T + \frac{EA}{2l}\int_0^l \left( \frac{\partial Z}{\partial x}\right)^2 dx\right]{{{\partial ^2}Z(x,t) } \over {\partial {x^2}}} = 0.
\end{multline}
Here, $Z(x, t)$ is the out-of-plane flexural displacement of the beam at axial position $x$ and time $t$ [Fig. \ref{exposition_linear}(a)]; $l\times w \times h$ are the nominal dimensions along $x$, $y$ and $z$, $A = wh$ is the cross-sectional area, and $I = wh^3/12$ is the area moment of inertia of the beam; the beam has  Young's modulus $E$ and density $\rho$. The first term in Eq. (\ref{beam}) is the inertial term, the second term is the rigidity term, and the third term is the tension term. The first component of the tension term in the square brackets is the intrinsic  tension $F_T$, and the second term corresponds to the additional tension due to  stretching. Ignoring the nonlinear term for a moment and   non-dimensionalizing Eq. (\ref{beam}), we obtain the nondimensional tension parameter,\cite{Barbish2022, ari2020nanomechanical}
\begin{equation}
    U={F_T \over 2EI/l^2},
\end{equation}
which  represents the ratio of the axial load on the beam to its rigidity. The system approaches  an Euler-Bernoulli beam  for $U\ll 1$ and a string for $U \gg 1$.

\begin{table*}
%\vspace{10pt}
\caption{Experimental parameters for the first 11 modes of two NEMS resonators.}
\label{experimental}
\renewcommand{\arraystretch}{1.1}
\newcolumntype{Y}{>{\centering\arraybackslash}X}
\begin{tabularx}{\textwidth}{Y|YYYY|YYYY}
\hline\hline
 & \multicolumn{4}{c|}{$l\times w \times h = \rm 50~\mu m \times 900~nm \times 100~nm$, $U = 4233$} & \multicolumn{4}{c}{$l\times w \times h = \rm 30~\mu m \times 900~nm \times 100~nm$, $U = 214$}\\
 \hline
 Mode& $\omega_n/2\pi$ & $Q_n$ & $k_n$ & $\alpha_n$ & $\omega_n/2\pi$ & $Q_n$ & $k_n$ & $\alpha_n$\\
 & $(\rm{MHz})$ &  & $(\rm N/m)$ & $(\rm MHz^2/nm^2)$ & $(\rm{MHz})$ &  & $(\rm N/m)$ & $(\rm MHz^2/nm^2)$\\
 \hline
1 & 4.954 & $2.0\times 10^4 $& 9.05& $5.69 \times 10^{-4}$ & 3.488 & $9.3 \times 10^3 $& 2.41& $3.18 \times 10^{-3}$\\
2 & 9.927& $1.7\times 10^4$  & 30.2& $5.95 \times 10^{-3}$& 7.169   & $7.0\times 10^3$& 9.96& $4.62 \times 10^{-2}$\\
3 & 14.83& $1.5\times 10^4$  & 67.5& $2.15 \times 10^{-2}$& 11.32   & $6.1\times 10^3$& 18.7& $1.46 \times 10^{-1}$\\
4 & 19.74& $1.3\times 10^4$    & 91.5& $8.71 \times 10^{-2}$& 15.85   & $4.7\times 10^3$& 31.5& $4.35 \times 10^{-1}$\\
5 & 24.83& $9.8\times 10^3$   & 159& $2.68 \times 10^{-1}$& 20.66   & $3.5\times 10^3$& 55.5& $9.21 \times 10^{-1}$\\
6 & 30.08   & $8.5\times 10^3$   & 248& $6.55 \times 10^{-1}$& 26.01  & $2.9\times 10^3$& 99.7& $2.35$\\
7 & 35.15   & $ 6.5\times 10^3 $   & 320& $1.33$& 32.00   & $2.5\times 10^3 $& 196& $3.91$\\
8 & 40.22   & $ 4.0\times 10^3 $   & 440& $1.92$& 38.89& $1.2\times 10^3 $& 246& $7.54$\\  
9 & 45.52   & $ 3.2\times 10^3 $   & 642& $3.13$& 46.21& $ 1.2\times 10^3 $& 403& $2.23 \times 10^{1}$\\
10 & 50.78   & $ 2.2\times 10^3 $   & 739& $4.90$& 53.92& $ 9.1 \times 10^2 $& 516& $3.23 \times 10^{1}$\\ 
11 & 56.06   & $ 2.3\times 10^3 $   & 845& $6.58$& 62.60& $7.8 \times 10^2 $& 719& $9.91 \times 10^{1}$\\
\hline\hline
\end{tabularx}
\end{table*}

In the limit $U \gg 1$, we neglect the bending term to obtain  the (nonlinear) string equation
\begin{equation}\label{string}
    \rho A{{{\partial ^2} Z(x,t)} \over {\partial {t^2}}} - \left[ F_T + \frac{EA}{2l}\int_0^l \left( \frac{\partial Z}{\partial x}\right)^2 dx\right]{{{\partial ^2}Z(x,t) } \over {\partial {x^2}}} = 0,
\end{equation}
with boundary conditions $Z(0,t)=Z(l,t)=0$ and eigenfunctions $\phi_n(x) = \sin{\left(\frac{n\pi x}{l}\right)}$.  Expanding the solution $Z(x,t)$ in terms of the orthogonal $\phi_n(x)$ and integrating over the length of the beam~\cite{postma2005dynamic, lifshitz2008nonlinear} yields a Duffing equation for the time dependent amplitude $z_n(t)$ of the different modes.
Including a lumped drive force of amplitude $F_n$ at frequency $\omega \approx \omega_n$ as well as a small phenomenological damping term,  we arrive at \cite{kozinsky2007basins,postma2005dynamic,lifshitz2008nonlinear} 
\begin{equation}\label{eqn: duffing}
{\ddot{z}_n} + {{{\omega _n}} \over {{Q_n}}}{\dot{z}_n} + {{\omega_n}^2}{z_n} + {\alpha _n}{z_n}^3 = {F_n \over m_n} \cos{\omega t}.
\end{equation}
This eigenfunction expansion yields the usual expression for the eigenfrequency $\omega_n =  \frac{n \pi}{l} \sqrt{\frac{F_T}{\rho A}}$; the modal spring constant (referred to an antinode\cite{gress2023multimode}) and the  modal mass can  be found as    $k_n= \frac{\pi^2 F_T}{2l}n^2$  and   $m_n = k_n/{\omega_n}^2=\frac{\rho l A}{2}$, respectively.\cite{Barbish2022,hauer2013general} Most importantly, the  modal Duffing constant $\alpha_n$  emerges as\cite{lifshitz2008nonlinear}
\begin{equation}\label{alphan}
    \alpha_n = \left(\frac{E\pi^4}{4\rho l^4}\right)n^4.
\end{equation} 
With the above choice of the eigenfunctions, $z_n(t)$ corresponds to the actual (peak) displacement of the string at its antinodes. We note that, even for large $U$, the string approximation is expected to break down for large $n$, where bending  becomes important.

Returning to Eq. (\ref{eqn: duffing}) above, standard steps\cite{landau2013course, nayfeh2008nonlinear,lifshitz2008nonlinear,{kozinsky2007basins,postma2005dynamic}} lead to an expression for  the frequency of the peak $\omega_{p,n}$ as a function of the peak amplitude $z_{p, n}$ and other modal parameters:
\begin{equation}\label{eqn:backbone}
    \omega_{p,n} = \omega_n + \frac{3}{8}\left(\frac{{z_{p, n}}^2}{\omega_n}\right)\alpha_n.
\end{equation}
Eq. (\ref{eqn:backbone}) is  parabolic in $z_{p, n}$  and describes the backbone curve,  \cite{kozinsky2007basins, nayfeh2008nonlinear} which follows the peak as the amplitude increases.

The critical amplitude $z_{c,n}$ can be determined from Eq. (\ref{eqn: duffing}) by performing a multiple time-scale analysis,\cite{postma2005dynamic,nayfeh2008nonlinear} which yields
\begin{equation}\label{eqn:critical-experiment}
    {z_{c,n}} = \sqrt {{{8\sqrt 3 } \over 9}{{{\omega _n}^2} \over {{Q_n}{\alpha _n}}}}.
\end{equation}
Substituting for the string values of $\omega_n$ and $\alpha_n$ in Eq. (\ref{eqn:critical-experiment}) and realizing that ${U}= \left(\frac{l}{h}\right)^2\left ({\frac{6F_T}{EA}}\right )$, we arrive at\cite{lifshitz2008nonlinear}
\begin{equation}\label{critical-theory}
    z_{c,n} = \left(\frac{4h}{3^{5/4}\pi}\sqrt{\frac{U}{Q_n}}\right)\frac{1}{n}.
\end{equation}
The tension parameter $U$, when considered  in Eq.~(\ref{critical-theory}), represents the ratio of the intrinsic tension to the elasticity induced tension due to the stretching of the beam.

Our experimental study is based on two silicon nitride (SiN) NEMS doubly clamped beam resonators with different intrinsic tension values. These two resonators are fabricated on two different wafers, one with higher stress than the other. A colored scanning electron microscope (SEM) image of the high-tension  NEMS beam is shown in Fig. \ref{exposition_linear}(a). Here, the beam lies along the $x$ axis, and out-of-plane vibrations are actuated in the $z$ direction. The nominal dimensions of the beam are $l\times w \times h \approx \rm 50~\mu m \times 900~nm \times 100~nm$. The suspended region is shown in bright green  and is clamped to the substrate underneath on either side. Gold nanoresistors are patterned on top of the beam near the clamps for electrothermal actuation. Nanoresistors have a width of $120 \rm~nm$ and thickness of $135 \rm ~nm$. All dimensions of the low-tension beam are the same, except the nanoresistor thickness is $60 \rm ~nm$ and $l \approx \rm 30~\mu m$.   To measure linear and nonlinear parameters of each device, we apply a sinusoidal current to the electrothermal actuators\cite{ti2022dynamics} and sweep the frequency from low to high around each eigenfrequency. We detect the NEMS motion interferometrically\cite{ti2022dynamics} at the forcing frequency using a lock-in amplifier.  By gradually increasing the actuation current, we drive each eigenmode from its linear to nonlinear regime.   For each eigenmode,  both linear and nonlinear  measurements are taken at an antinode  closest to the center of the beam. All the experiments are performed in a high vacuum  chamber.

Electrothermal actuation is based on Joule heating of the beam around the nanoresistor by an input current. \cite{ma2023electrothermal} Coupled with the  piezoresistance\cite{ti2021frequency} and the temperature-dependent resistivity of the gold nanoresistor,  Joule heating can result in mechanical forces and responses at the harmonics of the drive frequency. The lock-in amplifier filters out the higher harmonics, and the mixed down signal components are estimated to be negligible. We have also estimated the nonlinearity of the optical interferometer, and found it to be a     small but observable source of extrinsic nonlinearity. Finally, nonlinear dissipation is not appreciable in our system (see supplementary materials for details). 

\begin{figure}[h!]
    \includegraphics{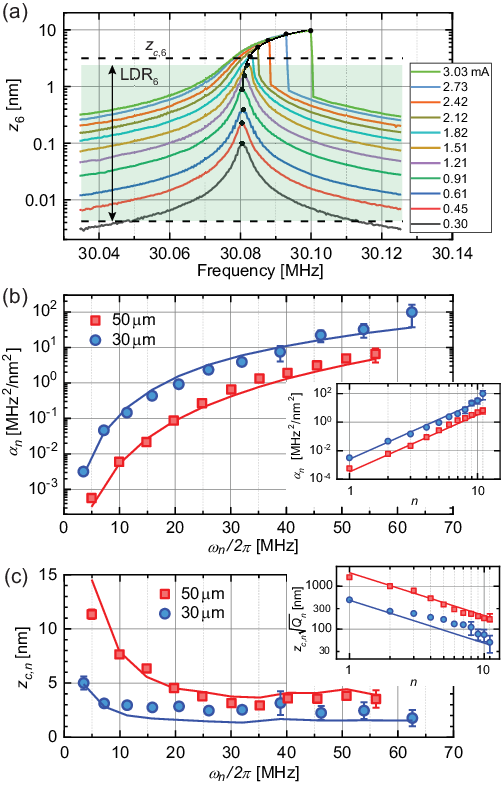}
    \caption{(a) Frequency-response curves (peak) of the $6^{\rm th}$ mode of the $50$-$\rm {{\mu} m}$ beam for different drive current values (rms). The upper dashed line shows the critical amplitude $z_{c, 6}$. The lower dashed line is the rms thermal amplitude $\sqrt{k_BT/k_6}$. The green-shaded area corresponds to the LDR. The Duffing constant $\alpha_6$ is extracted from a parabolic fit to the backbone curve. (b) Duffing constants $\alpha_n$ as a function of eigenfrequency shown on a semilogarithmic plot. Inset shows $\alpha_n$ as a function of $n$ on a double-logarithmic plot, revealing the trend $\alpha_n \propto n^4$. (c) Critical amplitudes $z_{c, n}$ (peak) as a function of eigenfrequency plotted on linear axes. The theoretical curves are non-smooth due to experimental $Q_n$. Inset shows $z_{c, n}\sqrt{Q_n}$ as a function of $n$ on a double-logarithmic plot, revealing the trend $z_{c, n}\sqrt{Q_n} \propto {1}/{n}$.}
    \label{nonlinear_measurements}
\end{figure}

We first measure the linear modal parameters of the NEMS beams [Figs. \ref{exposition_linear}(b)-\ref{exposition_linear}(d)], including the eigenfrequencies $\omega_n/2\pi$, spring constants $k_n$, quality factors $Q_n$, and compute the intrinsic tensions $F_T$. Fig. \ref{exposition_linear}(b) shows $\omega_n/2\pi$ as a function of mode number $n$ in a semilogarithmic plot. Here, the red and blue symbols correspond to the experimental values for the frequencies of the $50$-$\rm {{\mu} m}$ and $30$-$\rm {{\mu} m}$ beams, respectively, and the continuous lines correspond to the values computed using tensioned beam theory.\cite{ari2020nanomechanical,Barbish2022,gress2023multimode}  The lower inset shows the normalized frequencies, $\omega_n/\omega_1$, on a linear plot and reveals a strong string-like behavior for the $50$-$\rm {{\mu} m}$ beam. We find $F_T$ by matching analytical eigenfrequencies $\omega_n/2\pi$ with experiments. For material properties of SiN, we use a Young's modulus of $E = 250~\rm GPa$, \cite{villanueva2014evidence,klass2022determining} a Poisson's ratio of $\nu = 0.23$, and a mass density $\rho = 3000~\rm kg/m^3$. We find $F_T = 63.5~\rm \mu N$ for the $50$-$\rm {{\mu} m}$ beam and $F_T = 8.9~\rm \mu N$ for the $30$-$\rm {{\mu} m}$ beam.  The respective $U$ values for the two beams are $4233$ and $214$. The upper inset in Fig. \ref{exposition_linear}(b) shows the displacement profile for a typical eigenmode, the $7^{\rm th}$ eigenmode for the $50$-$\rm {{\mu} m}$ beam.

Fig. \ref{exposition_linear}(c) shows  $k_n$ as a function of $n$ in a semilogarithmic plot. To find the experimental $k_n$, we measure the power spectral density (PSD) of thermal fluctuations for each mode at a modal antinode closest to the center of the beam. We first find the mean-squared fluctuation amplitude by numerically integrating the modal PSD, we then calculate the spring constants from the equipartition of energy.\cite{gress2023multimode} A representative PSD for the $7^{\rm th}$ eigenmode of the $50$-$\rm {{\mu} m}$ beam is shown in the inset, where the continuous line is a Lorentzian fit to the peak. The continuous lines in the main figure are found from tensioned beam theory using mechanical properties of the beam and $F_T$.\cite{Barbish2022,gress2023multimode} Finally, Fig.  \ref{exposition_linear}(d) shows the experimental quality factors $Q_n$, which  are determined from linear time-domain ringdowns and frequency-domain Lorentzian fits\cite{ti2021frequency} to the resonator  response (see supplementary materials for details). The inset shows the same data on a semilogarithmic plot. $Q_n$ decreases monotonically with increasing mode number $n$. The observed decrease in $Q_n$ as a function of $n$ is consistent with dissipation dilution,\cite{unterreithmeier2010damping,cross2001elastic,yu2012control,fedorov2019generalized} and our overall dissipation is likely dominated by the gold film near the anchors.\cite{villanueva2014evidence}

Next, we show how to extract the nonlinear Duffing constant $\alpha_n$ and the critical amplitude $z_{c,n}$ from a measurement of the NEMS modal response as a function of frequency. Fig. \ref{nonlinear_measurements}(a) shows the $6^{\rm th}$ mode resonance of the 50-$\rm \mu m$ beam for different rms drive current levels using a logarithmic $y$ axis. The first four response curves at the lowest drive levels are linear. With increasing drive power, the response exhibits nonlinear stiffening: the frequency of the peak increases with the drive, and, at the largest drives, the response makes a downward jump as the response is no longer a singled-valued function of frequency.\cite{landau2013course, nayfeh2008nonlinear}  We extract $\alpha_n$ by fitting experimental $z_{p,n}$ values at  larger drives to Eq. (\ref{eqn:backbone}).

Modal $\alpha_n$ values for both beams are shown as a function of $\omega_n/2\pi$ in Fig. \ref{nonlinear_measurements}(b). The $y$-axis  is  logarithmic  and has units of $\rm MHz^2/nm^2$ to yield reasonable numerical values. The inset is a  double-logarithmic plot of $\alpha_n$ \textit{vs.} $n$. The slope reveals that  $\alpha_n \propto n^4$. The continuous lines are from theory and are discussed below. All experimentally measured $\alpha_n$ along with linear modal parameters are listed in Table \ref{experimental}. 

Fig. \ref{nonlinear_measurements}(c) (symbols) shows experimental critical amplitudes $z_{c, n}$ as a function of frequency computed from the corresponding $\alpha_n$ using Eq. (\ref{eqn:critical-experiment}).  In the inset, we remove the $Q_n$ dependence of the data by plotting $z_{c,n} \sqrt{Q_n}$ as a function of $n$ in a double-logarithmic plot. The data scale as $z_{c, n}\sqrt{Q_n} \propto \frac{1}{n}$. The continuous theory lines are also discussed below. It is important to note that $z_{c,n}$ that are directly determined from modal response curves, \textit{i.e.}, by finding the amplitude at which the curve first attains infinite slope, agree with $z_{c,n}$ found using Eq. (\ref{eqn:critical-experiment}) to within $\sim10 \%$.

{For both beams $U \gg 1$, and it seems reasonable to use the string approximation, \textit{i.e.}, Eqs. (\ref{alphan}) and (\ref{critical-theory}), in order to determine the theoretical values for $\alpha_n$ and $z_{c,n}$. To calculate the theoretical curves in Figs. 2(b) and 2(c), we use the nominal dimensions of the beams, mechanical properties of SiN, and experimental values for $U$ and $Q_n$ in Eqs. (\ref{alphan}) and (\ref{critical-theory}), respectively. The  use of experimentally measured $Q_n$ in Eq. (\ref{critical-theory})  results in the non-smooth theoretical curves in Fig. \ref{nonlinear_measurements}(c). The agreement between experimental and theoretical $\alpha_n$ in Fig. \ref{nonlinear_measurements}(b) is very good for both beams. In Fig. \ref{nonlinear_measurements}(c),  the experimental   $z_{c,n}$ data deviate from  theory for the  $30$-$\rm {{\mu} m}$ beam at some of the large $n$ values. The reason for this can be traced to the string approximation and its neglect of bending contributions which  become increasingly important at larger $n$. Eq.~(\ref{eqn:critical-experiment}) indicates that any errors in $\omega_n$ will be propagated to $z_{c,n}$. Indeed, Fig. \ref{exposition_linear}(b) inset shows that $\omega_n$ for the $30$-$\rm {{\mu} m}$ beam deviate  from the string frequencies (continuous red curve) for high $n$.

Finally, we discuss the LDR of these nanomechanical resonators. We modify  Eq. (\ref{eqn:dr_initial}) to redefine the LDR of mode $n$ as
\begin{equation}\label{dr}
\mathrm{LDR}_n  = 20~{\log}\left(\frac{0.745z_{c,n}^{(rms)}}{\sqrt{k_BT/k_n}}\right),
\end{equation}
where the entire noise bandwidth is taken as the detection bandwidth.  Eq. (\ref{dr}) gives the value of $\mathrm{LDR}_n$ referred to an antinode. Returning to Fig. \ref{nonlinear_measurements}(a), the dashed line on the bottom shows the rms thermal amplitude, $\sqrt{k_BT/k_6}$. The modal critical amplitude $z_{c, 6}$ is indicated by the upper dashed line, which corresponds to the first displacement amplitude with infinite slope.\cite{landau2013course, nayfeh2008nonlinear} The green-shaded area corresponds to the LDR, with the upper end of the LDR slightly below $z_{c, 6}$. 

We compute the experimental $\mathrm{LDR}_n$ using experimental  values for $z_{c, n}$ [symbols in Fig. \ref{nonlinear_measurements}(c)] and  $k_n$ [symbols in Fig. \ref{exposition_linear}(c)]. For the theoretical curves, we use the theoretical   $z_{c, n}$ [Eq. (\ref{critical-theory})], in which $U$ and $Q_n$ are from experiments, and theoretical $k_n$ [continuous lines in Fig. \ref{exposition_linear}(c)]. Fig. \ref{modal_dr_v2}(a) shows the experimental and theoretical $\mathrm{LDR}_n$ as a function of  $n$ for the first 11 modes of the two NEMS resonators. Note, again, that the theoretical curve is non-smooth due to the experimentally-measured $Q_n$ values. Both experimental and theoretical LDR show a weak dependence on $n$.

\begin{figure}
    \includegraphics{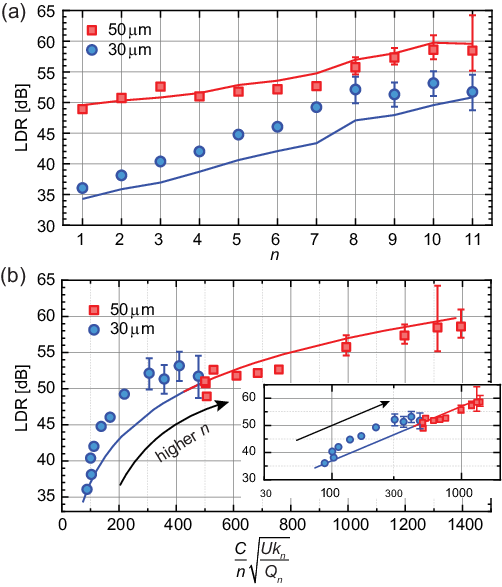}
    \caption{(a) Experimental and theoretical LDR for 11 eigenmodes of two NEMS resonators computed using Eq. (\ref{dr}). The theoretical curves are non-smooth due to experimental $Q_n$. (b) Collapsed LDR data. Inset shows the same data on a double-logarithmic plot, revealing a slope of $1$. }
    \label{modal_dr_v2}
\end{figure}

To better understand the dependence of LDR on physical parameters, we analyze the data in Fig. \ref{modal_dr_v2}(a) based on  Eqs. (\ref{critical-theory}) and (\ref{dr}).  We notice that, for the string approximation, the data should  only be a function of $\frac{\mathcal{C}}{n}\sqrt{\frac{Uk_n}{Q_n}}$, where $\mathcal{C} \approx\frac{0.24h}{\sqrt{k_BT}}$ is the same constant for both beams. This suggests that all data can be plotted as a function of $\frac{\mathcal{C}}{n}\sqrt{\frac{Uk_n}{Q_n}}$. To determine the new $x$ coordinate for each data point in Fig. \ref{modal_dr_v2}(a), we compute the corresponding value of $\frac{\mathcal{C}}{n}\sqrt{\frac{Uk_n}{Q_n}}$  based on all the experimental values. We then plot the data point against this $x$ value. The result is the plot in Fig. \ref{modal_dr_v2}(b). The theory curve is the smooth function $20\log \left(\frac{\mathcal{C}}{n}\sqrt{\frac{Uk_n}{Q_n}}\right)$. The inset shows the same data using a logarithmic $x$ axis and reveals a slope of $1$. The agreement between experimental and theoretical LDR for the $50$-$\rm {{\mu} m}$ beam is excellent, while the theory underpredicts experimental LDR of the $30$-$\rm {{\mu} m}$ beam. The arrows indicate the direction of increasing $n$ for the experimental data.  

At a first glance,  LDR  in Fig. \ref{modal_dr_v2}(a) increases  with $n$ and  $U$. However, $k_n$ and $Q_n$ both depend on $U$ and $n$. More insight can be gained into the  trends by turning to the $U\gg1$ limit. To this end, we use the string $k_n$ expressed in terms of $U$ as $k_n \approx \frac{\pi^2 U E I}{l^3}  n^2$ to find ${\rm LDR}_n \approx 20 \log \left(0.8h \sqrt{{EI} \over {l^3k_B T}}\right) + 20 \log \left( \frac{U}{\sqrt{Q_n}}\right)$, 
both referred to an antinode. For a given string,  ${\rm LDR}_n$  increases with increasing $U$, because  $z_{c,n}$ grows and  the thermal fluctuation amplitude decreases with $U$---extending the LDR on both ends. In the string approximation, ${\rm LDR}_n$ does not explicitly depend upon $n$  but only through $Q_n$. Intrinsic (\textit{in vacuo}) $Q_n$ tends to decrease\cite{ti2021frequency,bargatin2007efficient, yu2012control} with $n$, whereas fluidic $Q_n$ may increase\cite{gress2023multimode,kara2017generalized} with $n$. Comparison between different structures made of different materials is also possible using our formulas and requires  knowledge of $E$ in addition to nominal dimensions, $U$ and $Q_n$.

Our aim here has been to uncover the scaling of $\alpha_n$, $z_{c,n}$, and ${\rm LDR}_n$ across eigenmodes in a nanomechanical resonator. To this end, we have formulated our theory based on the string equation [Eq. (\ref{string})], which yielded insightful closed-form analytical expressions.  We attribute the disagreements between theory and experiment for our lower tension beam to the string approximation and will resolve this in future work by employing the tensioned beam theory. Finally, the frequency  resolution, and in particular, mass sensitivity of our NEMS resonators should increase with $n$, given that LDR increases but the active mass tends to stay unchanged with $n$.

\section*{supplementary material}
See the supplementary material for details of computing the experimental values of $Q_n$ and $F_T$, the theoretical values for $k_n$ and $\omega_n$, as well as the error sources in the system. Error sources include intermodal force coupling, transduction nonlinearities, thermally-induced frequency shifts, and nonlinear damping.

\begin{acknowledgments}
We acknowledge support from the US NSF (Nos. CMMI 2001403, CMMI 1934271, CMMI 2001559, and CMMI 1934370). M. Ma acknowledges support from Boston University Nanotechnology Innovation Center BUnano Cross-Disciplinary Fellowship. We thank H. Gress for the SEM image of the NEMS device.
\end{acknowledgments}

\section*{author declarations}
The authors have no conflict to disclose.

\bibliographystyle{apsrev4-2}
\bibliography{Reference.bib}

\end{document}

\title{Supplementary Material for ``Mode-Dependent Scaling of Nonlinearity and Linear Dynamic Range in a NEMS Resonator"}

\newcommand{\BU}{Department of Mechanical Engineering, Division of Materials Science and Engineering, and the Photonics Center, Boston University, Boston, Massachusetts 02215, USA}

\newcommand{\VT}{Department of Mechanical Engineering, Virginia Tech, Blacksburg, Virginia 24061, United States}

\author{M. Ma}
\affiliation{\BU}

\author{N. Welles}
\affiliation{\VT}

\author{O. Svitelskiy}
\affiliation{Department of Physics, Gordon College, Wenham, Massachusetts 01984, USA}

\author{C. Yanik}
\affiliation{SUNUM, Nanotechnology Research and Application Center, Sabanci University, Istanbul, 34956, Turkey}

\author{I. I. Kaya}
\affiliation{SUNUM, Nanotechnology Research and Application Center, Sabanci University, Istanbul, 34956, Turkey}
\affiliation{Faculty of Engineering and Natural Sciences, Sabanci University, Istanbul, 34956, Turkey}

\author{M. S. Hanay}
\affiliation{Department of Mechanical Engineering, Bilkent University, Ankara, 06800, Turkey}
\affiliation{National Nanotechnology Research Center (UNAM), Bilkent University, Ankara, 06800, Turkey}

\author{M. R. Paul}
\affiliation{\VT}

\author{K. L. Ekinci}
\email[Electronic mail: ]{ekinci@bu.edu}
\affiliation{\BU}

\date{\today}

\maketitle

\tableofcontents

\newpage

\widetext

\setcounter{equation}{0}
\setcounter{figure}{0}
\setcounter{table}{0}
\setcounter{page}{1}
\makeatletter
\renewcommand{\theequation}{S\arabic{equation}}
\renewcommand{\thefigure}{S\arabic{figure}}
\renewcommand{\bibnumfmt}[1]{[#1]}
\renewcommand{\citenumfont}[1]{#1}

\maketitle

\section{Experimental Details}

\subsection{Actuation and Detection}

In all experiments, we apply a sinusoidal current around half the value of the eigenfrequency, $\omega_n \over 4\pi$, to one or both electrothermal actuators. In the case of  driving with both actuators, the two supplied current signals have the same frequency and power, and are in-phase for driving odd modes and out-of-phase for even modes. The drive configuration does not affect the dynamical response or the spatial profile of eigenmodes due to the very high $Q_n$ values \cite{ti2022dynamics}. Joule heating generates temperature oscillations in  the actuator at twice the input current frequency. Owing to the mismatch between the thermal expansion coefficients  of the gold film and the silicon nitride beam, a bending moment ensues \cite{ma2023electrothermal}, driving out-of-plane (along $z$) flexural oscillations of the resonator around the eigenfrequency, $\omega_n \over 2\pi$. The harmonically-driven dynamics of the NEMS resonator, \textit{i.e.}, the rms amplitude and phase of its oscillations, is measured using a homodyne optical interferometer. The frequency domain measurements are performed using a  lock-in amplifier. For time domain ring-down   measurements, the NEMS is driven on resonance electrothermally by a burst signal and its motion is detected on a digital oscilloscope.

\subsection{Quality Factors}\label{SIsubsection:qn}

For the $30$-$\rm {{\mu} m}$ beam, we extract the modal quality factors $Q_n$ by fitting experimental linear modal response curves to Lorentzian peak functions
\begin{equation}\label{eqn:lorentz-amplitude}
    z_{n}(\omega) = \frac{F_n/m_n}{\sqrt{({\omega_n}^2-\omega^2)^2+(\frac{\omega_n\omega}{Q_n})^2}}.
\end{equation}
Here, we use the same notation as in the main text: $F_n/m_n$ is the lumped modal force normalized by the modal mass; $\omega_n/2\pi$ is the eigenfrequency; and $\omega/2\pi$ is the drive frequency. Our reported $Q_n$ values are obtained by averaging   several linear fits at different drives, and the error bars are the standard deviations, as shown in Fig. 1(d) in the main text. 
\begin{figure*}[htbp]
   \includegraphics[width=6in]{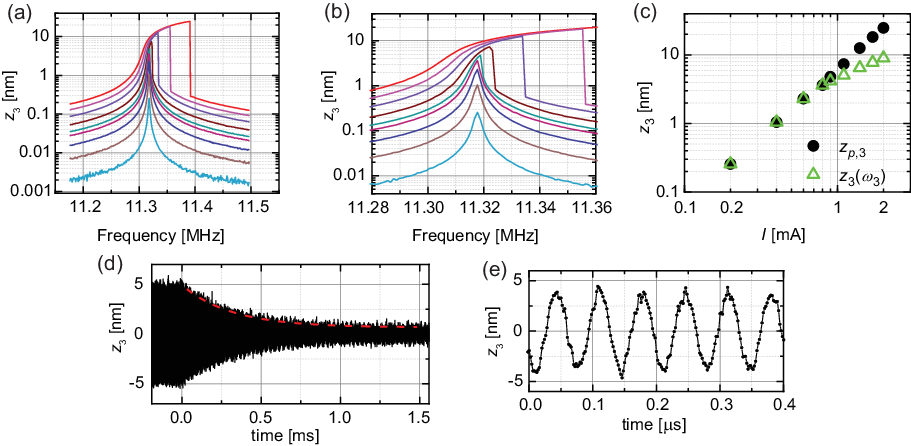}
    \caption{(a) Amplitude response (peak) curves as a function of frequency for the $3^{\rm rd}$ mode of the $30$-$\rm {{\mu} m}$ beam shown on a semilogarithmic plot. (b) Close-up view of linear response peaks. (c) Peak response amplitudes (black circles) and response amplitudes at the linear resonance frequency (green triangles) as a function of the drive current plotted on double-logarithmic axes.  (d) Time-domain ringdown response (peak) for the $3^{\rm rd}$ mode of the $50$-$\rm {{\mu} m}$ beam. Drive force is turned off  at time $t = 0$. An exponential fit to the upper envelope of the ringdown is shown by the red dashed line. (e) Close-up view of the time-domain signal.}
    \label{SIfig:1}
\end{figure*}

Fig. \ref{SIfig:1}(a) shows a set of  driven  response curves as a function of frequency  on semilogarithmic axes. Fig. \ref{SIfig:1}(b) shows a close-up view of the peaks of the linear response curves. The peak amplitude for both linear and nonlinear response increases linearly with the drive power, \textit{i.e.}, $z_{p,n}\propto I^2$ [Fig. \ref{SIfig:1}(c) black circles], where $I$ is the peak amplitude of the drive current. Once nonlinearity becomes appreciable, however, the amplitude at the linear eigenfrequency, $z_3(\omega_3)$, is no longer the peak amplitude [Fig. \ref{SIfig:1}(c) green triangles].
%which are the bottom four curves in Fig. \ref{SIfig:1}(a).

For the $50$-$\rm {{\mu} m}$ beam, we determine $Q_n$ from  modal ringdowns when the NEMS resonator is driven to a high, but still linear, amplitude response close to the eigenfrequency. The linear modal resonator response during a time-domain ringdown is described by
\begin{equation}\label{eqn:ringdown}
    z_n(t) = z_{p,n} e^{-{\pi f_n t}/{Q_n}} \cos (2\pi f_n + \varphi) + z_{N,n},
\end{equation}
where $z_{p,n}$ is the peak modal antinode displacement, $\varphi$ is an arbitrary phase, and $z_{N,n}$ is the thermal noise. To compute $Q_n$, we fit the upper envelope of the ringdown to Eq. (\ref{eqn:ringdown}) [Fig. \ref{SIfig:1}(d) red dashed line]. We compute $Q_n$ using both Lorentzian fits and the ringdown method, and the results agree within $5\%$ for most eigenmodes. We use ringdown-based $Q_n$ for the $50$-$\rm {{\mu} m}$ beam due to the lower standard deviations across  measurements.

We show a sample ringdown  measurement for the $3^{\rm rd}$ mode in Fig. \ref{SIfig:1}(d). We drive the resonator with a burst signal for $2 \times 10^4$ cycles and a burst period long enough to observe the ringdown. In Fig. \ref{SIfig:1}(d), the drive signal stops at time $t = 0$, and the exponential decay fit to the upper envelope of the ringdown signal is shown by the red dashed line. The rms thermal noise value is $z_{N, 3} \approx 0.5~\rm nm$. In Fig. \ref{SIfig:1}(e), we show a close-up view of the time-domain signal to reveal the oscillations at the eigenfrequency.

\section{Computing Theoretical Parameters from Tensioned Beam Theory}

Here, we compute the intrinsic tension force as well as the theoretical values of eigenfrequencies and spring constants using the Euler-Bernoulli beam theory with tension \cite{barbish2022dynamics, gress2023multimode}.

We first estimate the magnitude of the tension force $F_T$ by comparing the eigenfrequencies obtained from experiments, $\omega_n/2\pi$, with theoretical predictions, $\omega_n^{(t)}/2\pi$. To determine $\omega_n^{(t)}/2\pi$ for a beam under tension with the same nominal dimensions as our devices, we first define the nondimensional eigenfrequency as
\begin{equation}
 C_n=\left(\frac{\omega_n^{(t)}}{\beta/l^2}\right)^{1/2}.  
\label{eq:C_n}
\end{equation}
Here, $\beta=(EI/\mu)^{1/2}$. We determine the mass per unit length from $\mu=\rho wh$ using $\rho \approx 3000~\rm kg/m^3$ \cite{ti2021frequency}. The area moment of inertia is determined from $I=\frac{wh^3}{12}$ based on nominal beam dimensions, and the Young's modulus is taken as $E=250~\rm GPa$. The characteristic equation \cite{bokaian1990natural}
\begin{multline}
   {C_n}^2+U\sinh{\left[\left(U+\sqrt{U^2+{C_n}^4}\right)^{1/2}\right]}\sin{\left[\left(-U+\sqrt{U^2+{C_n}^4}\right)^{1/2}\right]}\\-{C_n}^2\cosh{\left[\left(U+\sqrt{U^2+{C_n}^4}\right)^{1/2}\right]}\cos{\left[\left(-U+\sqrt{U^2+{C_n}^4}\right)^{1/2}\right]}=0
\label{eq:C_n_U}
\end{multline}
relates $C_n$ to the nondimensional tension parameter $U=\frac{F_T}{2EI/l^2}$. Thus,  $\omega_n^{(t)}/2\pi$ is determined by finding the roots of Eq.~(\ref{eq:C_n_U}) for a given value of $F_T$ using nominal dimensions, $\rho$ and $E$.  

For each beam, we solve for  $\omega_n^{(t)}/2\pi$ as the value of $F_T$ is swept, and then compute the cumulative error 
\begin{equation}\label{SIeqn:error}
    \varepsilon(F_T) = \sum_{n=1}^{11} \frac{(\omega_n^{(t)} - \omega_n)^2}{{\omega_n}^2}
\end{equation}
between $\omega_n$ and $\omega_n^{(t)}$, where $n$ encompasses the first eleven modes. We assume that the minimum  $\varepsilon$ is achieved for the experimental value of $F_T$. %Because all beams are on the same chip, we average the values of $F_T$ for each of our three beams. 
We find that $F_T = 63.5~\rm \mu N$ for the $50$-$\rm {{\mu} m}$ beam and $F_T = 8.9~\rm \mu N$ for the $30$-$\rm {{\mu} m}$ beam. The respective $U$ values for the two beams are $4233$ and $214$. In summary, the value for $\omega_n^{(t)}$ plotted in Fig. 1(b) in main text is based on this experimental value of $F_T$, nominal beam dimensions, $\rho=3000~\rm kg/m^3$ and $E=250~\rm GPa$. 

We find the theoretical spring constants $k_n(x_n)$  for the tensioned beam at an antinode $x_n$ from $k_n(x_n) = \frac{m{\omega_n}^2}{{\phi_n(x_n)}^2}$ \cite{gress2023multimode}. Here, we first calculate the eigenfunctions $\phi_n(x)$ for each tensioned beam using the corresponding value of $F_T$ and $\omega_n^{(t)}/ 2\pi$ \cite{barbish2022dynamics}; we then use the nominal mass $m=\mu l$ and evaluate $\phi_n(x_n)$ to determine $k_n(x_n)$. Our experimental spring constants match theoretical predictions over several orders of magnitude, as shown in Fig. 1(c) in the main text.

\section{Error Sources}\label{SIsection:extrinsic-nonlin}

\subsection{Intermodal Drive Force Coupling}\label{SIsubsection:intermodal_force}

The nonlinear components of the actuation force for a given eigenmode are unlikely to excite the other eigenmodes of the beam \textit{resonantly} because of the frequency mismatch and the large $Q_n$. For the high-tension beam, for example, the $n^{\rm{th}}$ eigenfrequency is close to  an integer multiple of the fundamental eigenfrequency, $\omega_n \approx n \omega_1$, and this could potentially be an issue. We estimate this intermodal  coupling between the fundamental and the second eigenmodes since this would be the strongest. In this case, $\frac{\omega_2 - 2\omega_1}{\omega_2/Q_2}  \approx 33 \gg 1$ and thus the force is \textit{very far} from the resonance condition. The coupling should be even worse between other modes given the frequency mismatch will be larger.  As a result,  sweeping the frequency of the actuation force  near an eigenfrequency should only  result in \textit{off-resonance} excitations of higher modes, and the ensuing  response will not be  appreciable.

\subsection{Transducer Nonlinearities}

In a \textit{harmonically-driven} nonlinear mechanical resonator, mechanical responses will be generated at many different frequencies  due to mechanical mixing. The general trend is that, if the nonlinearity is small, terms at higher frequencies become progressively smaller.  In our NEMS device,  the  drive force comes from an electrothermal actuator, which is inherently nonlinear---in addition to the mechanical nonlinearity. In our devices, the electrothermal actuator is a gold nanoresistor fabricated around the anchor of the beam. An input oscillatory current   heats up the nanoresistor, leading to thermoelastic oscillations in the beam \cite{ma2023electrothermal}. The electrothermal force acting on the beam and the beam displacement amplitude  are proportional to the oscillatory temperature $\Delta T$, which, in turn, is proportional to the power dissipated in the gold nanoresistor \cite{ma2023electrothermal}. In addition,  the resistance of the nanoresistor is time-dependent:  small oscillatory terms in resistance emerge due to the stretching  of the nanoresistor during beam oscillations, \textit{i.e.}, the metallic piezoresistive effect \cite{ti2021frequency}, and due to the temperature-dependent resistivity of gold.

We  start with a NEMS resonator that is oscillating nonlinearly due to an electrothermal force at $\omega_n$. This force results from  a constant-amplitude sinusoidal voltage, $V(t)=V^{(0)} {\cos\frac{\omega_n t}{2}}$, that is applied to the nanoresistor on the NEMS \footnote{We have reported the excitation current values in the main text and Sections \ref{SIsubsection:qn} and \ref{SIsubsection:thermal_shift}---instead of the voltage. In the experiments, we indeed apply a constant voltage $V^{(0)}$ to the electrothermal transducer, but it is  more informative to report the current value, $V^{(0)} \over R^{(0)}$, so that the effects of the contact and embedding resistances are removed from the problem. Given that ${R^{(1)}\over R^{(0)}} \ll 1$ (see Subsections \ref{Sisub2section:piezo} and \ref{Sisub2section:heating}), the current is also practically constant.}. Here, $V^{(0)}$ is the peak amplitude of the applied voltage. The response amplitude of the NEMS can be expressed as $z_n(t)=z_n^{(1)} \cos \left({\omega _n}t+ \theta _n^{(1)}\right)+ z_n^{(2)}\cos \left({2\omega _n}t+ \theta _n^{(2)}\right)+...$, where the superscript indicates the term order, and $\theta _n^{(n)}$ corresponds to the displacement phase at frequency $n\omega _n$. Each term is smaller than the one preceding it, as usual \cite{landau2013course}. We assume that the resistance of the nanoresistor can be expressed  as  
\begin{equation}\label{SIeqn:resistance_general}
    R(t)-R^{(0)} \approx f(z_n),
\end{equation}
where $R^{(0)}$ is the unperturbed value of the resistance. The  justification for Eq. (\ref{SIeqn:resistance_general}) is physical: i) as the beam oscillates, its resistance is modulated due to its piezoresistance; and ii)  the oscillatory temperature $\Delta T$ of the beam, which is proportional to the beam amplitude, modulates the  resistance due to the temperature dependence of resistivity (see below). If we now expand $f(z_n)$ in a power series and rearrange, we can cast Eq. (\ref{SIeqn:resistance_general}) into the form
\begin{equation}\label{SIeqn:resistance_nonlinear}
    R(t)=R^{(0)}\left[1+ {R^{(1)}\over R^{(0)}}{\cos{\omega_n}t}+  {R^{(2)} \over R^{(0)}}{\cos\left({2\omega_n}t+\varphi_n^{(2)}\right)}+...\right],
\end{equation}
where the superscript indicates the  order of the term and $\varphi_n^{(n)}$ corresponds to the  phase. Then, the current in the nanoresistor will also oscillate at many frequencies given that $I(t)={V(t) \over R(t)} \approx {V^{(0)} \over R^{(0)}} {\cos\frac{\omega_n t}{2}} \left[1- {R^{(1)}\over R^{(0)}}{\cos{\omega_n}t}- ...\right]$. The temperature and hence the force on the beam will be proportional to the power dissipated,  $F_n(t) \propto \Delta T(t) \propto {I(t)  V(t)}$. Ignoring the dc components, we arrive at 
\begin{equation}\label{SIeqn:force_nonlinear_approx}
    F_n(t) \approx F_n {\cos{\omega_n}t} \left[{{1} - {{R^{(1)}\over R^{(0)}}{\cos{\omega_n}t} -  {R^{(2)}\over R^{(0)}}{\cos\left({2\omega_n}t+\varphi_n^{(2)}\right)} -... }}\right].
\end{equation}
Eq. (\ref{SIeqn:force_nonlinear_approx}) shows that the electrothermal actuation force and hence the beam displacement will have many components at integer multiples of $\omega_n$, \textit{i.e.}, $m\omega_n$. It is also clear from Eq. (\ref{SIeqn:force_nonlinear_approx}) that some components of the response, \textit{e.g.},  $ {R^{(2)}\over R^{(0)}}\,{\cos\left({2\omega_n}t+\varphi_n^{(2)}\right)}$ will be mixed down to $\omega_n$.

In our measurements, the lock-in amplifier is sharply tuned to the  frequency $\omega_n$ of the force applied to the beam.  Thus, the lock-in detects  the  amplitude and phase of the oscillations of the beam at exactly $\omega_n$.  In other words, only the response at  $\omega_n$ is detected, and the response at other frequencies, including higher harmonics, are rejected. The discussion above in Section \ref{SIsubsection:intermodal_force} suggests that the harmonics of the force will not resonantly excite the other eigenmodes.  In addition, electrical mixing between the input and output transducers is not an issue here since we use optical detection, which has negligible coupling to the input transducer, \textit{i.e.}, the gold nanoresistor. However, harmonic signals that are mixed down to  the exact detection frequency $\omega_n$ will introduce errors to the measurement.

In order to provide some insight into the above-discussed  error, we will  estimate the magnitude of  the   $\frac{R^{(1)}}{R^{(0)}}$ term coming from  metallic piezoresistance and  heating. The expectation is that all higher order terms, \textit{e.g.}, 
${{{F_n}{R^{(2)}}} \over {{R^{(0)}}}}\cos {\omega _n}t \,\cos \left( {2{\omega _n}t + \varphi _n^{(2)}} \right)$, will be even  smaller. Therefore, if the $\frac{R^{(1)}}{R^{(0)}}$ term is small, which is the case as shown below, we will conclude that our measurements are not affected by these  mixed-down signals.

\subsubsection{Metallic Piezoresistive Effect}\label{Sisub2section:piezo}

Out-of-plane beam oscillations stretch a NEMS transducer along its length. We express the relative change in resistance due to geometry \cite{li2007ultra} as
\begin{equation}\label{SIeqn:geometric-piezo}
    {{R_p}^{(1)} \over R^{(0)}} = \gamma \epsilon(z_n).
\end{equation}
Here, $\gamma$ is the gauge factor, and $\epsilon(z_n)$ is the average strain in the gold nanoactuator caused by the beam displacement amplitude $z_n$. The strain in the nanoactuator is
\begin{equation}\label{Sieqn:strain}
    \epsilon(z_n) = \frac{\Delta l_e}{l_e},
\end{equation}
where $l_e$ is the length of the nanoactuator. We estimate the nanoactuator elongation $\Delta l_e$ from the beam elongation due to an out-of-plane displacement $z_n$, 
\begin{equation}\label{SIeqn: pifagor}
    \frac{dl}{dx} \approx \sqrt{1+\left(z_n\frac{\partial \phi_n}{\partial x}\right)^2},
\end{equation}
where $dl$ is the beam elongation, and $\frac{\partial \phi_n}{\partial x}$ is the spatial derivative of an eigenmode. After binomial approximation, we find the elongation by integrating over the length of the nanoresistor:
\begin{equation}\label{SIeqn:resistor-elongation}
    \Delta l_e = \frac{{z_n}^2}{2} \int_{0}^{l_e} \left(\frac{\partial \phi_n}{\partial x}\right)^2 \,dx.
\end{equation}
Taking $z_n$ as the highest experimental antinode displacement (peak value) of mode $n$, $l_e \approx 600~\rm nm$, $\gamma \approx 10$ \cite{ti2021frequency} and the string eigenmodes as $\phi_n = \sin{(\frac{n\pi x}{l})}$, we compute the largest value of ${{R_p}^{(1)} \over R^{(0)}}$ due to the metallic piezoresistive effect across $11$ eigenmodes of both beams and find
\begin{equation}\label{SIeqn:geometric-piezo_max}
    {{R_p}^{(1)} \over R^{(0)}} = \frac{\gamma {z_n}^2 \int_{0}^{l_e} \left(\frac{\partial \phi_n}{\partial x}\right)^2 \,dx}{2l_e} \lesssim 10^{-3}.
\end{equation}
These results are consistent with measurements in \cite{ti2021frequency}.

\subsubsection{Resistance Change due to Heating}\label{Sisub2section:heating}

For a drive current at frequency $\frac{\omega_n}{2}$, the leading order temperature oscillations in the beam are at $\omega_n$. This  results in resistance oscillations at the same frequency $\omega_n$ \cite{lu20013omega,zhou2007determination}. The relative magnitude of oscillatory resistance is proportional to the oscillatory temperature
\begin{equation}\label{SIeqn:3omega}
    {{R_t}^{(1)} \over R^{(0)}} = k_t \Delta T \lesssim 10^{-2}.
\end{equation}
Here, $k_t = 0.0035~\rm K^{-1}$ is the temperature coefficient of resistance for gold and $\Delta T$ is the oscillatory temperature on the nanoactuator. 
We compute the value of maximum temperature rise as $\Delta T \approx 10~\rm K$ using FEM modeling based on the highest supplied drive current in experiments  \cite{ma2023electrothermal}.

\subsubsection{Optical Interferometer}

On the optical detection side, interferometer nonlinearities arise when beam oscillations are large compared to the optical wavelength \cite{wagner1990optical, dolleman2017amplitude}. We quantify the optical transduction error by computing the responsivity curves \cite{lifshitz2008nonlinear} for the fundamental modes of our NEMS devices, where we achieve the highest displacement amplitudes.

In a homodyne Michelson interferometer, one coverts mechanical displacement into a phase difference between the reference and the object laser beams, and the phase difference results in a measurable change in optical intensity on the photodetector. One can express the detected optical intensity $D$ as a function of object displacement $z$ as \cite{wagner1990optical, dolleman2017amplitude}
\begin{equation}\label{SIeqn:intereferometer-nonlinear}
    D = ({a_o}^2 + {a_r}^2)\left[ 1 + 2 \frac{a_r a_o}{{a_o}^2 + {a_r}^2}\cos\left(k (d_r - d_o) + 2kz\right) \right].
\end{equation}
Here, $a_o$ and $a_r$ are electric field amplitudes of the object and the reference laser beams, respectively, $d_r - d_o$ is the path difference between the two laser beams, $\lambda$ is the laser wavelength, and $k = 2\pi/\lambda$ is the wavenumber. For $z \ll \lambda$, one can use the small-amplitude approximation \cite{wagner1990optical}
\begin{equation}\label{SIeqn: interferemeter-linear}
    D \approx ({a_o}^2 + {a_r}^2)\left[ 1 + 2 \frac{a_r a_o}{{a_o}^2 + {a_r}^2}\cos(k [d_r - d_o]) - 2 \frac{a_r a_o}{{a_o}^2 + {a_r}^2}2kz\sin(k [d_r - d_o]) \right].
\end{equation}
Note the dependence of $D \propto z$ after linearization. For our optical setup, the path difference is actively stabilized using a PID controller at $d_r - d_o = \lambda/4$ for optimal sensitivity \cite{wagner1990optical}, $\lambda = 633~\rm nm$, and the largest measured peak displacement in our experiments is $\approx 40~\rm nm$. Using the linear approximation for optical calibration [Eq. (\ref{SIeqn: interferemeter-linear})], we get an error of $\lesssim 5\%$. 

\begin{figure*}[h]
\centering
   \includegraphics[width = 6.7in]{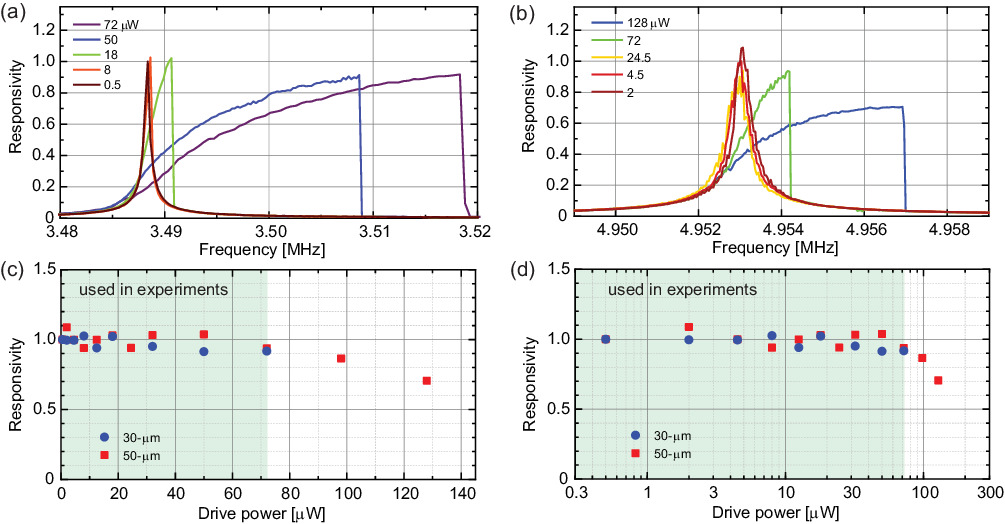}
    \caption{Responsivity curves of the fundamental modes for (a) the $30$-$\rm {{\mu} m}$ beam and (b) the $50$-$\rm {{\mu} m}$ beam. (c) Responsivity values as a function of drive power for both beams. Values used in experiments are highlighted. Data points at higher drives are for illustrating the optical  nonlinearities. (d) Data in (c) shown in a semilogarithmic plot.}
    \label{fig:responsivity_optical}
\end{figure*}

To experimentally characterize the optical transduction nonlinearity, we have calculated the responsivity curves for the fundamental mode of our beams---as these are the modes that achieve the largest amplitudes. The responsivity is defined as the response amplitude scaled by the drive power  \cite{lifshitz2008nonlinear}, \textit{i.e.}, $\frac{z_{p,n}}{I^2R^{(0)}}$, where $I$ is the drive current. We normalize all responsivity values to the average responsivity of linear responses. Thus, in the case of a perfectly linear transducer, the responsivity should be unity for all drive power levels. We show our experimental responsivity curves for the $30$-$\rm {{\mu} m}$ beam [Fig. \ref{fig:responsivity_optical}(a)] and $50$-$\rm {{\mu} m}$ beam [Fig. \ref{fig:responsivity_optical}(b)]. The responsivity values for each drive power are summarized in Fig. \ref{fig:responsivity_optical}(c) for both beams. The same data is shown on a semilogarithmic plot in Fig. \ref{fig:responsivity_optical}(d).  We show the data points used for our experiments under the green-shaded area. Two additional measurements at higher drive powers were taken for the $50$-$\rm {{\mu} m}$ beam to show the error due to optical interferometry. In our experiments, the error value for the highest displacement is $\approx 8 \%$, in good agreement with Eq. (\ref{SIeqn: interferemeter-linear}). 

\subsection{Thermal Frequency Shifts}\label{SIsubsection:thermal_shift}

Electrothermal actuation creates a temperature increase in the nanoresistor region, which can cause a downshift in the eigenfrequency. Higher modes are mechanically stiffer; thus, one needs more actuation power and higher temperatures to achieve nonlinear response. In our devices, non-negligible thermal frequency shifts emerge for mode $n = 11$ of the $50$-$\rm {{\mu} m}$ beam and modes $n \geq 8$ of the $30$-$\rm {{\mu} m}$ beam.

We remove the thermal frequency shift as follows. We perform a frequency sweep with a weak drive signal so that the resonator response is linear. At the same time, we supply an off-resonance harmonic signal to heat the electrode. We track the linear resonance change at different power levels of the off-resonance signal, which allows us to obtain a curve for the thermal frequency shift as a function of power. Using this thermal frequency shift, we offset the frequency in our nonlinearity measurements for similar power levels. 

\begin{figure*}[h]
   \includegraphics[width=4in]{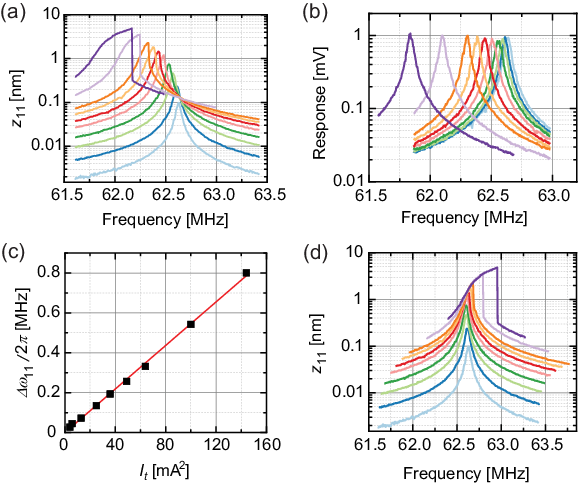}
    \caption{(a) Uncorrected experimental response curves at different drives as a function of frequency for the $11^{\rm th}$ mode of the $30$-$\rm {{\mu} m}$ beam. Thermal frequency downshifts overshadow the nonlinearity induced upshifts. (b) Response curves as a function of frequency at different off-resonance heating powers. Here, the eigenfrequency shifts are due to thermal effects only. (c) Thermal eigenfrequency shift $\Delta \omega_{11}/2\pi$ as function of the squared current amplitude of the off-resonance signal ${I_t}^2$. Symbols represent experimental data extracted from (b), and reveal a linear trend, $\Delta \omega_{11}/2\pi \propto {I_t}^2$. Solid line is a linear fit. (d) Experimental response curves as a function of frequency after correction.}
    \label{SIfig:3}
\end{figure*}

We show an example of the process of removing the thermal frequency shift for the $11^{\rm th}$ mode of the $30$-$\rm {{\mu} m}$ beam [Fig. \ref{SIfig:3}]. Fig. \ref{SIfig:3}(a) shows the uncorrected response as a function of frequency. Here, the upshift of the frequency of the peak $\omega_{p,11}/2\pi$ due to nonlinearity is overshadowed by the thermal frequency downshifts. To remove the thermal frequency shifts, we drive the resonator with a small current signal of $I_r = 1~\rm mA$ (peak) and sweep the frequency near the linear eigenfrequency $\omega_{11}/2\pi$. In addition, we supply a harmonic signal $I_{t}$ at a constant frequency away from $\omega_{11}/2\pi$. Sweeping the amplitude of $I_{t}$ results in the resonance curves shown in Fig. \ref{SIfig:3}(b). We plot the thermal frequency shift $\Delta \omega_{11}/2\pi$ as a function of the squared current of the off-resonance signal ${I_t}^2$, which shows a linear relationship between the drive power and the thermal frequency shift $\Delta \omega_{11}/2\pi$ [Fig. \ref{SIfig:3}(c)]. Finally, we remove the thermal frequency shifts by offsetting the frequency in our amplitude response curves based on the drive power and the linear fit shown in Fig. \ref{SIfig:3}(c). Fig. \ref{SIfig:3}(d) shows the response curves as a function of frequency with removed thermal frequency shifts. 

\subsection{Nonlinear Damping}

\begin{figure*}[h]
\centering
   \includegraphics[width = 6.6in]{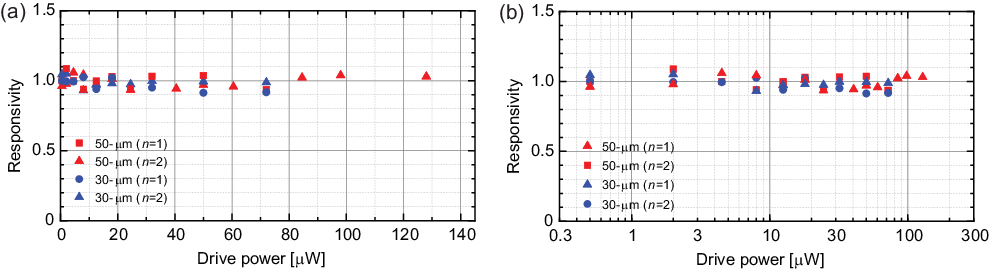}
    \caption{(a) Responsivity values as a function of drive power levels used in experiments for the first two modes of both beams. (b) Data in (a) shown in a semilogarithmic plot.}
    \label{fig:responsivity_damping}
\end{figure*}

We make an attempt to quantify the nonlinear damping in our beams by computing the responsivity curves \cite{lifshitz2008nonlinear}. We show the responsivity of both beams for modes $n = 1$ and $n=2$ in linear [Fig. \ref{fig:responsivity_damping}(a)] and semilogarithmic [Fig. \ref{fig:responsivity_damping}(b)] plots. All responsivity values are normalized by the average responsivity of the linear response. For $n=1$, the responsivity decreases at higher drive levels. However, this trend has a significant contribution from the optical transduction nonlinearity, which, similar to nonlinear damping effects, becomes appreciable at higher displacement amplitudes. As a result, it is difficult to decouple the individual contribution of nonlinear damping. 

We estimate the responsivity decrease due to nonlinear damping to be $\lesssim 3\%$ for $n=1$ by subtracting the theoretical error value [Eq. \ref{SIeqn: interferemeter-linear}] for optical interferometry. For $n=2$, the responsivity is nearly constant at higher drive powers, and thus nonlinear damping should be negligible for $n\ge 2$.

% \bibliographystyle{apsrev4-2}
% \bibliography{Reference.bib}
%\printbibliography
